\documentclass[showpacs,preprintnumbers,amsmath,amssymb]{revtex4}

\usepackage{graphicx}
\usepackage{dcolumn}
\usepackage{bm}

\newcommand{\bq}{\begin{equation}}
\newcommand{\eq}{\end{equation}}
\newcommand{\bqa}{\begin{eqnarray}}
\newcommand{\eqa}{\end{eqnarray}}
\newcommand{\nn}{\nonumber \\}

\def\be     {\begin{equation}}
\def\ee     {\end{equation}}
\def\bea        {\begin{eqnarray}}
\def\eea        {\end{eqnarray}}
\def\bnn    {\begin{eqnarray*}}
\def\enn    {\end{eqnarray*}}

\begin{document}

\title{$z = 3$ antiferromagnetic quantum criticality and emergence of fermionized skyrmions}
\author{Ki-Seok Kim}
\affiliation{ Department of Physics, POSTECH, Hyoja-dong, Namgu,
Pohang, Gyeongbuk 790-784, Korea \\ Institute of Edge of
Theoretical Science (IES), Hogil Kim Memorial building 5th floor,
POSTECH, Hyoja-dong, Namgu, Pohang, Gyeongbuk 790-784, Korea }
\date{\today}

\begin{abstract}
Hertz-Moriya-Millis theory with dynamical critical exponent $z =
2$ has been proposed to describe antiferromagnetic quantum
criticality in itinerant electron systems. In this study we show
that the dynamical critical exponent changes from $z = 2$ to $z =
3$ at low temperatures, which results from effective long-range
interactions between spin fluctuations, generated by Fermi-surface
fluctuations beyond the Eliashberg framework. We claim that the
underlying physics for the $z = 3$ antiferromagnetic quantum
criticality is the emergence of fermionized skyrmion excitations
at low energies, which form a Fermi surface, referred as a
skyrmion liquid, where the interplay between itinerant electrons
and skyrmions is argued to allow fermionized skyrmions. We
construct a dual field theory in terms of skyrmion excitations,
and obtain the $z = 3$ antiferromagnetic quantum criticality. This
demonstration suggests the expected but nontrivial consistency
between weak-coupling and strong-coupling theories.
\end{abstract}


\maketitle

\section{Introduction}

At present, it is almost impossible to understand non-perturbative
physics based on the weak coupling approach, where physically
relevant quantum corrections should be taken into account up to an
infinite order. Actually, we do not know how to describe various
strong coupling phenomena such as the Kondo effect, Mott
transition, Anderson localization, and etc., starting from the
Fermi liquid theory and performing the diagrammatic analysis.
Dualities between weak-coupling and strong-coupling theories have
been playing key roles in figuring out non-perturbative phenomena,
where topologically nontrivial excitations instead of original
variables are introduced to describe strong coupling physics.
Domain wall excitations appear to describe quantum criticality in
the dual field theory of Ising model, identified with fermions,
where the transformation from the Ising variable to the fermion
field is highly nonlocal and nonlinear
\cite{Ising_Duality_Review}. Unfortunately, duality
transformations are not well defined in most cases. It is also
almost impossible to derive the corresponding dual theory. Even if
a dual field theory can be constructed, it is hard to prove the
equivalence of physics at quantum criticality between the original
and dual field theories \cite{Tesanovic_vs_Kleinert}.

However, the situation becomes better in one dimension
\cite{Bosonization_textbook}. Starting from an interacting
one-dimensional model, one can perform the weak coupling analysis
to find self-consistent equations for self-energy corrections of
electrons, renormalized interactions, and vertex corrections.
Resorting to the Ward identity, one obtains asymptotically exact
effective interactions, which allow us to find the electron
Green's function. It turns out not to have poles but to show two
branch cuts, implying the spin-charge separation and identified
with Luttinger liquid. This diagrammatic analysis can be
reformulated in terms of topologically nontrivial excitations
called spinons and holons, which carry only spin and charge
quantum numbers, respectively. This duality transformation is
called bosonization, where duality equations are highly nonlinear
and nonlocal as the case of Ising model. Resorting to such dual
variables, one can construct a dual field theory in terms of
topological excitations, which results in asymptotically the same
Green's function as that of the diagrammatic approach.

In this study we address essentially the same issue on
non-perturbative dynamics in the Fermi-surface problem. Before
2009, it has been believed that the $1/N_{\sigma}$ expansion,
where $N_{\sigma}$ denotes the spin degeneracy to extend from
$\uparrow, \downarrow$ to $1, 2, ..., N_{\sigma}$, is a
controllable method in understanding quantum criticality of
itinerant electron systems, sometimes referred as the Eliashberg
framework or a self-consistent RPA (random phase approximation)
theory \cite{Pepin_Chubukov}. This belief in suspense has been
broken in 2009, where all planar diagrams turn out to be in the
``same" order even for the $1/N_{\sigma}$ expansion
\cite{Beyond_Eliashberg_Lee}. This conclusion means that vertex
corrections associated with planar diagrams should be introduced
up to an infinite order in order to obtain reliable critical
exponents near quantum criticality \cite{Beyond_Eliashberg_Max}.

We investigate two-dimensional antiferromagnetic quantum
criticality in itinerant electron systems. Hertz-Moriya-Millis
theory with dynamical critical exponent $z = 2$ has been proposed
to describe the antiferromagnetic quantum criticality. However, we
show that the dynamical critical exponent changes from $z = 2$ to
$z = 3$ at low temperatures, which results from effective
long-range interactions between spin fluctuations, generated by
Fermi-surface fluctuations beyond the Eliashberg framework. We
argue that the underlying physics for the $z = 3$
antiferromagnetic quantum criticality is the emergence of
fermionized skyrmion excitations at low energies, which form a
Fermi surface, referred as a skyrmion liquid, where the interplay
between itinerant electrons and skyrmions is argued to allow
fermionized skyrmions. We construct a dual field theory in terms
of skyrmion excitations, and obtain the $z = 3$ antiferromagnetic
quantum criticality. This demonstration suggests the expected but
nontrivial consistency between weak-coupling and strong-coupling
theories.

\section{Hertz-Moriya-Millis-Chubukov theory revisited}

\subsection{Spin-fermion model}

A ``standard" model for the spin-density-wave quantum criticality
is the spin-fermion model, given by \bqa && Z =
\int{Dc_{n\sigma}D\boldsymbol{\phi}} e^{- \int_{0}^{\beta}{d\tau}
\int d^{d} \boldsymbol{r} {\cal L}} , \nn && {\cal L} =
c_{n\sigma}^{\dagger} (\partial_{\tau} - \mu_{c}) c_{n\sigma} +
\frac{1}{2m_{c}} |(\boldsymbol{\nabla} - i \boldsymbol{A})
c_{n\sigma}|^{2} + \lambda_{\phi} e^{i
\boldsymbol{Q}\cdot\boldsymbol{r}} \boldsymbol{\phi} \cdot
c_{n\alpha}^{\dagger} \boldsymbol{\sigma}_{\alpha\beta} c_{n\beta}
+ \boldsymbol{\phi} (-\partial_{\tau}^{2} - v_{\phi}^{2}
\boldsymbol{\nabla}^{2} + \xi^{-2}) \boldsymbol{\phi} , \eqa where
$c_{n\sigma}$ is an electron field with spin $\sigma = \uparrow,
\downarrow$ and $n = 1, ..., N$, and $\boldsymbol{\phi}$ is a
spin-density field with three components. $\mu_{c}$ is an electron
chemical potential, and $m_{c}$ is an electron band mass.
$\boldsymbol{A}$ is an electro-magnetic vector potential field,
externally applied and not considered in this study. $v_{\phi}$ is
a spin-wave velocity, and $\xi$ is a spin-spin correlation length.
The antiferromagnetic quantum critical point is defined as the
renormalized correlation length diverges, i.e., $\xi
\longrightarrow \infty$. $\lambda_{\phi}$ is a coupling constant
between spin fluctuations and itinerant electrons, and
$\boldsymbol{Q}$ is a nesting wave vector. Although we focus on
two-dimensional quantum criticality, we leave the dimension to be
represented by $d$.

\subsection{Eliashberg theory}

The spin-fermion model has been solved in the Eliashberg
framework, where only self-energy corrections of electrons and
spin fluctuations are taken into account self-consistently while
vertex corrections are neglected \cite{Chubukov_Review}.
Self-consistent equations are given by \bqa &&
\Sigma(\boldsymbol{k},i\omega) = 3 \lambda_{\phi}^{2}
\frac{1}{\beta} \sum_{i\Omega} \sum_{\boldsymbol{q}}
G(\boldsymbol{k}+\boldsymbol{q},i\omega+i\Omega)
\chi(\boldsymbol{q},i\Omega) , \nn && \Pi(\boldsymbol{q},i\Omega)
= - N \frac{3\lambda_{\phi}^{2}}{2}  \frac{1}{\beta}
\sum_{i\omega} \sum_{\boldsymbol{k}}
G(\boldsymbol{k}+\boldsymbol{q},i\omega+i\Omega)
G(\boldsymbol{k},i\omega) , \eqa where
$\Sigma(\boldsymbol{k},i\omega)$ and $\Pi(\boldsymbol{q},i\Omega)$
are self-energy corrections of electrons and spin fluctuations,
respectively, and $G(\boldsymbol{k},i\omega)$ and
$\chi(\boldsymbol{q},i\Omega)$ are renormalized electron Green's
function and spin susceptibility, given by Dyson equations \bqa &&
G(\boldsymbol{k},i\omega) = \frac{1}{i\omega + \mu_{c} -
\epsilon_{\boldsymbol{k}} - \Sigma(\boldsymbol{k},i\omega)} , \nn
&& \chi(\boldsymbol{q},i\Omega) = \frac{1}{\Omega^{2} +
v_{\phi}^{2} \boldsymbol{q}^{2} + \xi^{-2} +
\Pi(\boldsymbol{q},i\Omega)} , \eqa respectively. These equations
can be solved easily at the antiferromagnetic quantum critical
point, solutions of which are $\Pi(\boldsymbol{q},i\Omega) \propto
N \lambda_{\phi}^{2} v_{F}^{-1} |\Omega|$ for the spin-fluctuation
self-energy and $\Sigma(i\omega) \propto \lambda_{\phi}^{2} N_{F}
\sqrt{|\omega|}$ for the electron self-energy in two dimensions,
respectively, where $N_{F}$ is the density of states at the Fermi
energy. We note that the scaling relation between momentum and
frequency in the spin-fluctuation propagator changes from
$\Omega^{2} \sim v_{\phi}^{2} |\boldsymbol{q}|^{2}$ to $|\Omega|
\sim [v_{\phi}^{2} v_{F} / (N \lambda_{\phi}^{2})]
|\boldsymbol{q}|^{2}$ at low frequencies due to the
spin-fluctuation self-energy. As a result, we obtain the dynamical
critical exponent of $z = 2$ for this antiferromagnetic quantum
critical point.

\subsection{Hertz-Moriya-Millis-Chubukov theory}

The $z = 2$ quantum critical liquid state, described by the
Eliashberg theory for spin fluctuations, has been proposed to be
the zeroth-order mother state at the quantum critical point as the
Fermi liquid state away from quantum critical points
\cite{Pepin_Chubukov}. In order to go beyond the Eliashberg
framework, we integrate over electronic degrees of freedom, their
dynamics of which is given by the Eliashberg theory, and expand
the resulting logarithmic term up to the fourth order in spin
fluctuations. As a result, we obtain an effective field theory for
spin fluctuations, \bqa && {\cal S}_{eff} = \frac{1}{\beta}
\sum_{i\Omega} \int \frac{d^{d} \boldsymbol{q}}{(2\pi)^{d}}
\boldsymbol{\phi}_{r}(i\Omega,\boldsymbol{q}) \Bigl( \Omega^{2} +
v_{\phi}^{2} \boldsymbol{q}^{2} + \gamma_{\psi} |\Omega| +
\xi^{-2} \Bigr) \boldsymbol{\phi}_{r}(-i\Omega,-\boldsymbol{q})
\nn && + \frac{1}{\beta} \sum_{i\Omega} \int \frac{d^{d}
\boldsymbol{q}}{(2\pi)^{d}} \frac{1}{\beta} \sum_{i\nu} \int
\frac{d^{d} \boldsymbol{p}}{(2\pi)^{d}} \frac{1}{\beta}
\sum_{i\nu'} \int \frac{d^{d} \boldsymbol{p}'}{(2\pi)^{d}}
\lambda_{4}(i\Omega,\boldsymbol{q};i\nu,\boldsymbol{p};i\nu',\boldsymbol{p}')
\nn && \Bigl\{
[\boldsymbol{\phi}_{r}(i\nu+i\Omega/2,\boldsymbol{p}+\boldsymbol{q}/2)
\cdot
\boldsymbol{\phi}_{r}(-i\nu+i\Omega/2,-\boldsymbol{p}+\boldsymbol{q}/2)]
[\boldsymbol{\phi}_{r}(i\nu'-i\Omega/2,\boldsymbol{p}'-\boldsymbol{q}/2)
\cdot
\boldsymbol{\phi}_{r}(-i\nu'-i\Omega/2,-\boldsymbol{p}'-\boldsymbol{q}/2)
] \nn && +
[\boldsymbol{\phi}_{r}(i\nu+i\Omega/2,\boldsymbol{p}+\boldsymbol{q}/2)
\times
\boldsymbol{\phi}_{r}(-i\nu+i\Omega/2,-\boldsymbol{p}+\boldsymbol{q}/2)]
\cdot
[\boldsymbol{\phi}_{r}(i\nu'-i\Omega/2,\boldsymbol{p}'-\boldsymbol{q}/2)
\times
\boldsymbol{\phi}_{r}(-i\nu'-i\Omega/2,-\boldsymbol{p}'-\boldsymbol{q}/2)
] \Bigr\} , \nn \eqa sometimes referred as the
Hertz-Moriya-Millis-Chubukov theory. $\gamma_{\psi} = N
\lambda_{\phi}^{2} v_{F}^{-1}$ is the Landau damping coefficient,
given by Eq. (2) with Eq. (3), and $\xi$ is a renormalized
correlation length.
$\lambda_{4}(i\Omega,\boldsymbol{q};i\nu,\boldsymbol{p};i\nu',\boldsymbol{p}')$
is a renormalized interaction vertex with four legs, given by
\cite{Chubukov_Review} \bqa &&
\lambda_{4}(i\Omega,\boldsymbol{q};i\nu,\boldsymbol{p};i\nu',\boldsymbol{p}')
\nn && = \frac{1}{4 !} \Bigl\{ \frac{1}{\beta} \sum_{i\omega} \int
\frac{d^{d} \boldsymbol{k}}{(2\pi)^{d}} G(\boldsymbol{k},i\omega)
G(\boldsymbol{k} + \boldsymbol{p} - \boldsymbol{q}/2, i\omega + i
\nu - i \Omega/2) G(\boldsymbol{k} - \boldsymbol{q},i\omega -
i\Omega) G(\boldsymbol{k} - \boldsymbol{p}' - \boldsymbol{q}/2,
i\omega - i \nu' - i \Omega/2) \Bigr\} , \nn \eqa where
$G(\boldsymbol{k},i\omega)$ is the Eliashberg electron Green's
function, given by Eq. (3).

An essential aspect is that the renormalized vertex of
$\lambda_{4}(i\Omega,\boldsymbol{q};i\nu,\boldsymbol{p};i\nu',\boldsymbol{p}')$
depends on the momentum of $\boldsymbol{q}$ in a singular way,
giving rise to long range interactions effectively in real space.
More precisely, such long range interactions occur when the
frequency of $\Omega$ becomes finite. If one considers the limit
of $\Omega \longrightarrow 0$, this renormalized interaction
recovers a constant interaction-vertex, referred as the
Hertz-Moriya-Millis theory.

We investigate the role of such nonlocal correlations in
non-perturbative dynamics of spin fluctuations, which originate
from Fermi-surface fluctuations beyond the Eliashberg
approximation. Actually, their effects have been examined
perturbatively, where vertex corrections associated with the
nonlocal term are introduced into the Eliashberg spin
susceptibility \cite{Beyond_Eliashberg_Max}. The dynamical
critical exponent turns out to get a correction term, resulting in
$z \not= 2$. On the other hand, we perform the RPA analysis beyond
the previous study, where the renormalization for the
$\lambda_{4}(i\Omega,\boldsymbol{q};i\nu,\boldsymbol{p};i\nu',\boldsymbol{p}')$
vertex is taken into account. This non-perturbative analysis gives
rise to $z = 3$ antiferromagnetic quantum criticality.

\subsection{Molecular approximation}

In order to renormalize the interaction vertex of
$\lambda_{4}(i\Omega,\boldsymbol{q};i\nu,\boldsymbol{p};i\nu',\boldsymbol{p}')$,
it is necessary to simplify the nonlocal interaction term, but
keeping an essential feature. An idea is to introduce ``molecular"
field variables, which consist of composites of original spin
variables. Although this procedure can be regarded to be quite
conventional, our main approximation is to neglect their internal
structures and to keep their center-of-mass dynamics only.

Performing the Fourier transformation for Eq. (4), we start from
the Hertz-Moriya-Millis-Chubukov theory in real space \bqa &&
{\cal S}_{eff} = \int_{0}^{\beta} d \tau \int d^{d} \boldsymbol{r}
\boldsymbol{\phi}_{r}(\boldsymbol{r},\tau) \Bigl( -
\partial_{\tau}^{2} - v_{\phi}^{2} \boldsymbol{\nabla}_{\boldsymbol{r}}^{2} + \gamma_{\psi}
\sqrt{- \partial_{\tau}^{2}} + \xi^{-2} \Bigr)
\boldsymbol{\phi}_{r}(\boldsymbol{r},\tau) \nn && +
\int_{0}^{\beta} d \tau_1 \int d^{d} \boldsymbol{r}_1
\int_{0}^{\beta} d \tau_2 \int d^{d} \boldsymbol{r}_2
\int_{0}^{\beta} d \tau_3 \int d^{d} \boldsymbol{r}_3
\int_{0}^{\beta} d \tau_4 \int d^{d} \boldsymbol{r}_4 \nn &&
\lambda_{4}[\boldsymbol{r}_{2}-\boldsymbol{r}_{1},\tau_{2}-\tau_{1};
\boldsymbol{r}_{4}-\boldsymbol{r}_{3},\tau_{4}-\tau_{3};
(\boldsymbol{r}_{4}+\boldsymbol{r}_{3})/2-(\boldsymbol{r}_{2}+\boldsymbol{r}_{1})/2,(\tau_{4}+\tau_{3})/2-(\tau_{2}+\tau_{1})/2]
\nn && \Bigl\{ [\boldsymbol{\phi}_{r}(\boldsymbol{r}_{1},\tau_{1})
\cdot \boldsymbol{\phi}_{r}(\boldsymbol{r}_{2},\tau_{2}) ]
[\boldsymbol{\phi}_{r}(\boldsymbol{r}_{3},\tau_{3}) \cdot
\boldsymbol{\phi}_{r}(\boldsymbol{r}_{4},\tau_{4}) ] +
[\boldsymbol{\phi}_{r}(\boldsymbol{r}_{1},\tau_{1}) \times
\boldsymbol{\phi}_{r}(\boldsymbol{r}_{2},\tau_{2})] \cdot
[\boldsymbol{\phi}_{r}(\boldsymbol{r}_{3},\tau_{3}) \times
\boldsymbol{\phi}_{r}(\boldsymbol{r}_{4},\tau_{4}) ] \Bigr\} .
\eqa It is straightforward to rewrite this field theory in terms
of center-of-mass and relative coordinates, given by \bqa && {\cal
S}_{eff} = \int_{0}^{\beta} d \tau \int d^{d} \boldsymbol{r}
\boldsymbol{\phi}_{r}(\boldsymbol{r},\tau) \Bigl( -
\partial_{\tau}^{2} - v_{\phi}^{2} \boldsymbol{\nabla}_{\boldsymbol{r}}^{2} + \gamma_{\psi}
\sqrt{- \partial_{\tau}^{2}} + \xi^{-2} \Bigr)
\boldsymbol{\phi}_{r}(\boldsymbol{r},\tau) \nn && +
\int_{0}^{\beta} d \tau_1 \int d^{d} \boldsymbol{r}_1
\int_{0}^{\beta} d \tau_2 \int d^{d} \boldsymbol{r}_2
\int_{0}^{\beta} d \mathcal{T}_1 \int d^{d} \boldsymbol{R}_1
\int_{0}^{\beta} d \mathcal{T}_2 \int d^{d} \boldsymbol{R}_2
\lambda_{4}[\boldsymbol{r}_{1},\tau_{1};
\boldsymbol{r}_{2},\tau_{2};
\boldsymbol{R}_{2}-\boldsymbol{R}_{1},\mathcal{T}_{2}-\mathcal{T}_{1}]
\nn && \Bigl\{
[\boldsymbol{\phi}_{r}(\boldsymbol{R}_1+\boldsymbol{r}_1/2,\mathcal{T}_1+\tau_1/2)
\cdot
\boldsymbol{\phi}_{r}(\boldsymbol{R}_1-\boldsymbol{r}_1/2,\mathcal{T}_1-\tau_1/2)
]
[\boldsymbol{\phi}_{r}(\boldsymbol{R}_2+\boldsymbol{r}_2/2,\mathcal{T}_2+\tau_2/2)
\cdot
\boldsymbol{\phi}_{r}(\boldsymbol{R}_2-\boldsymbol{r}_2/2,\mathcal{T}_2-\tau_2/2)
] \nn && +
[\boldsymbol{\phi}_{r}(\boldsymbol{R}_1+\boldsymbol{r}_1/2,\mathcal{T}_1+\tau_1/2)
\times
\boldsymbol{\phi}_{r}(\boldsymbol{R}_1-\boldsymbol{r}_1/2,\mathcal{T}_1-\tau_1/2)
] \cdot
[\boldsymbol{\phi}_{r}(\boldsymbol{R}_2+\boldsymbol{r}_2/2,\mathcal{T}_2+\tau_2/2)
\times
\boldsymbol{\phi}_{r}(\boldsymbol{R}_2-\boldsymbol{r}_2/2,\mathcal{T}_2-\tau_2/2)
] \Bigr\} , \eqa where $\boldsymbol{R}_{1} =
\frac{\boldsymbol{r}_{1} + \boldsymbol{r}_{2}}{2}$,
$\mathcal{T}_{1} = \frac{\tau_{1} + \tau_{2}}{2}$ and
$\boldsymbol{R}_{2} = \frac{\boldsymbol{r}_{3} +
\boldsymbol{r}_{4}}{2}$, $\mathcal{T}_{2} = \frac{\tau_{3} +
\tau_{4}}{2}$ are center-of-mass coordinates and
$\boldsymbol{r}_{2} - \boldsymbol{r}_{1} \longrightarrow
\boldsymbol{r}_{1}$, $\tau_{2} - \tau_{1} \longrightarrow
\tau_{1}$ and $\boldsymbol{r}_{4} - \boldsymbol{r}_{3}
\longrightarrow \boldsymbol{r}_{2}$, $\tau_{4} - \tau_{3}
\longrightarrow \tau_{2}$ are relative coordinates.

We introduce molecular field variables, \bqa && \Bigl\langle
\boldsymbol{\phi}_{r}(\boldsymbol{R}_1+\boldsymbol{r}_1/2,\mathcal{T}_1+\tau_1/2)
\cdot
\boldsymbol{\phi}_{r}(\boldsymbol{R}_1-\boldsymbol{r}_1/2,\mathcal{T}_1-\tau_1/2)
\Bigr\rangle \longrightarrow
\mathcal{V}(\boldsymbol{r}_1,\tau_1;\boldsymbol{R}_1,\mathcal{T}_1)
, \nn && \Bigl\langle
\boldsymbol{\phi}_{r}(\boldsymbol{R}_1+\boldsymbol{r}_1/2,\mathcal{T}_1+\tau_1/2)
\times
\boldsymbol{\phi}_{r}(\boldsymbol{R}_1-\boldsymbol{r}_1/2,\mathcal{T}_1-\tau_1/2)
\Bigr\rangle \longrightarrow
\boldsymbol{\mathcal{X}}(\boldsymbol{r}_1,\tau_1;\boldsymbol{R}_1,\mathcal{T}_1)
, \eqa where
$\mathcal{V}(\boldsymbol{r}_1,\tau_1;\boldsymbol{R}_1,\mathcal{T}_1)$
represents valence bond fluctuations, and
$\boldsymbol{\mathcal{X}}(\boldsymbol{r}_1,\tau_1;\boldsymbol{R}_1,\mathcal{T}_1)$
expresses vector spin-chirality fluctuations.

Performing the Hubbard-Stratonovich transformation and doing the
Fourier transformation, we obtain \bqa && {\cal S}_{eff} =
\sum_{i\Omega} \sum_{\boldsymbol{q}}
\boldsymbol{\phi}_{r}(\boldsymbol{q},i\Omega) \Bigl( \Omega^{2} +
v_{\phi}^{2} \boldsymbol{q}^{2} + \gamma_{\psi} |\Omega| +
\xi^{-2} \Bigr) \boldsymbol{\phi}_{r}(-\boldsymbol{q},-i\Omega)
\nn && + \sum_{i\Omega} \sum_{\boldsymbol{q}} \frac{1}{\beta}
\sum_{i\Omega'} \sum_{\boldsymbol{q}'} \frac{1}{\beta}
\sum_{i\Omega_1} \sum_{\boldsymbol{q}_1} \frac{1}{\beta}
\sum_{i\Omega_1'} \sum_{\boldsymbol{q}_1'}
\lambda_{4}[-\boldsymbol{q}_{1},-i\Omega_1;-(\boldsymbol{q}+\boldsymbol{q}')/2,-(i\Omega+i\Omega')/2;\boldsymbol{q}_{1}',i\Omega_{1}']
\nn && \Bigl\{ \frac{1}{4}
\mathcal{V}(\boldsymbol{q}_1,i\Omega_1;\boldsymbol{q}_{1}',i\Omega_{1}')
\mathcal{V}((\boldsymbol{q}+\boldsymbol{q}')/2,(i\Omega+i\Omega')/2;-\boldsymbol{q}_{1}',-i\Omega_{1}')
- i
\mathcal{V}(\boldsymbol{q}_1,i\Omega_1;\boldsymbol{q}-\boldsymbol{q}',i\Omega-i\Omega')
[\boldsymbol{\phi}_{r}(\boldsymbol{q}',i\Omega') \cdot
\boldsymbol{\phi}_{r}(-\boldsymbol{q},-i\Omega) ] \nn && +
\frac{1}{4}
\boldsymbol{\mathcal{X}}(\boldsymbol{q}_1,i\Omega_1;\boldsymbol{q}_{1}',i\Omega_{1}')
\cdot
\boldsymbol{\mathcal{X}}((\boldsymbol{q}+\boldsymbol{q}')/2,(i\Omega+i\Omega')/2;-\boldsymbol{q}_{1}',-i\Omega_{1}')
- i
\boldsymbol{\mathcal{X}}(\boldsymbol{q}_1,i\Omega_1;\boldsymbol{q}-\boldsymbol{q}',i\Omega-i\Omega')
\cdot [\boldsymbol{\phi}_{r}(\boldsymbol{q}',i\Omega') \times
\boldsymbol{\phi}_{r}(-\boldsymbol{q},-i\Omega) ] \Bigr\} . \nn
\eqa Up to now, all procedures from Eq. (4) to Eq. (9) are exact.

An essential approximation is to neglect internal dynamics of
molecular field variables and to keep their center-of-mass
dynamics. Actually, this approximation has been rather
conventionally utilized because internal dynamics can be regarded
as fast degrees of freedom. However, we confess that the validity
of this approximation is not proven, particulary, at the quantum
critical point. If sizes of molecules become enhanced and the
probability for overlapping between them gets larger at the
quantum critical point, our approximation to view such molecules
as point particles will not be valid any more.

Performing the point-particle approximation for such composite
field variables, we obtain \bqa && {\cal S}_{eff} \approx
\sum_{i\Omega} \sum_{\boldsymbol{q}}
\boldsymbol{\phi}_{r}(\boldsymbol{q},i\Omega) \Bigl( \Omega^{2} +
v_{\phi}^{2} \boldsymbol{q}^{2} + \gamma_{\psi} |\Omega| +
\xi^{-2} \Bigr) \boldsymbol{\phi}_{r}(-\boldsymbol{q},-i\Omega)
\nn && + \sum_{i\Omega} \sum_{\boldsymbol{q}} \Bigl\{
\frac{\lambda_{4}(\boldsymbol{q},i\Omega)}{4}
\mathcal{V}(\boldsymbol{q},i\Omega)
\mathcal{V}(-\boldsymbol{q},-i\Omega) +
\frac{\lambda_{4}(\boldsymbol{q},i\Omega)}{4}
\boldsymbol{\mathcal{X}}(\boldsymbol{q},i\Omega) \cdot
\boldsymbol{\mathcal{X}}(-\boldsymbol{q},-i\Omega) \nn && - i
\Bigl( \frac{1}{\beta} \sum_{i\Omega_1'} \sum_{\boldsymbol{q}_1'}
\lambda_{4}(\boldsymbol{q}_{1}',i\Omega_{1}') \Bigr)
\frac{1}{\beta} \sum_{i\Omega'} \sum_{\boldsymbol{q}'}
\mathcal{V}(\boldsymbol{q}-\boldsymbol{q}',i\Omega-i\Omega')
[\boldsymbol{\phi}_{r}(\boldsymbol{q}',i\Omega') \cdot
\boldsymbol{\phi}_{r}(-\boldsymbol{q},-i\Omega) ] \nn && - i
\Bigl( \frac{1}{\beta} \sum_{i\Omega_1'} \sum_{\boldsymbol{q}_1'}
\lambda_{4}(\boldsymbol{q}_{1}',i\Omega_{1}') \Bigr)
\frac{1}{\beta} \sum_{i\Omega'} \sum_{\boldsymbol{q}'}
\boldsymbol{\mathcal{X}}(\boldsymbol{q}-\boldsymbol{q}',i\Omega-i\Omega')
\cdot [\boldsymbol{\phi}_{r}(\boldsymbol{q}',i\Omega') \times
\boldsymbol{\phi}_{r}(-\boldsymbol{q},-i\Omega) ] \Bigr\} . \eqa
Rescaling
$\mathcal{V}(\boldsymbol{q}-\boldsymbol{q}',i\Omega-i\Omega')
\longrightarrow
\mathcal{V}(\boldsymbol{q}-\boldsymbol{q}',i\Omega-i\Omega') /
\Bigl( \frac{1}{\beta} \sum_{i\Omega_1'} \sum_{\boldsymbol{q}_1'}
\lambda_{4}(\boldsymbol{q}_{1}',i\Omega_{1}') \Bigr)$,
$\boldsymbol{\mathcal{X}}(\boldsymbol{q}-\boldsymbol{q}',i\Omega-i\Omega')
\longrightarrow
\boldsymbol{\mathcal{X}}(\boldsymbol{q}-\boldsymbol{q}',i\Omega-i\Omega')
/ \Bigl( \frac{1}{\beta} \sum_{i\Omega_1'}
\sum_{\boldsymbol{q}_1'}
\lambda_{4}(\boldsymbol{q}_{1}',i\Omega_{1}') \Bigr)$, and
$\lambda_{4}(\boldsymbol{q},i\Omega) \longrightarrow \Bigl(
\frac{1}{\beta} \sum_{i\Omega_1'} \sum_{\boldsymbol{q}_1'}
\lambda_{4}(\boldsymbol{q}_{1}',i\Omega_{1}') \Bigr)^{2}
\lambda_{4}(\boldsymbol{q},i\Omega)$, we reach the following
expression \bqa && Z_{eff} = \int D
\boldsymbol{\phi}_{r}(\boldsymbol{q},i\Omega) D
\mathcal{V}(\boldsymbol{q},i\Omega) D
\boldsymbol{\mathcal{X}}(\boldsymbol{q},i\Omega) \exp\Big\{ -
\beta {\cal S}_{eff}
[\boldsymbol{\phi}_{r}(\boldsymbol{q},i\Omega),
\mathcal{V}(\boldsymbol{q},i\Omega),
\boldsymbol{\mathcal{X}}(\boldsymbol{q},i\Omega)] \Bigr\} , \nn &&
{\cal S}_{eff} = \sum_{i\Omega} \sum_{\boldsymbol{q}}
\boldsymbol{\phi}_{r}(\boldsymbol{q},i\Omega) \Bigl( \Omega^{2} +
v_{\phi}^{2} \boldsymbol{q}^{2} + \gamma_{\psi} |\Omega| +
\xi^{-2} \Bigr) \boldsymbol{\phi}_{r}(-\boldsymbol{q},-i\Omega)
\nn && + \sum_{i\Omega} \sum_{\boldsymbol{q}} \Bigl\{
\frac{\lambda_{4}(\boldsymbol{q},i\Omega)}{4}
\mathcal{V}(\boldsymbol{q},i\Omega)
\mathcal{V}(-\boldsymbol{q},-i\Omega) +
\frac{\lambda_{4}(\boldsymbol{q},i\Omega)}{4}
\boldsymbol{\mathcal{X}}(\boldsymbol{q},i\Omega) \cdot
\boldsymbol{\mathcal{X}}(-\boldsymbol{q},-i\Omega) \nn && - i
\frac{1}{\beta} \sum_{i\Omega'} \sum_{\boldsymbol{q}'}
\mathcal{V}(\boldsymbol{q}-\boldsymbol{q}',i\Omega-i\Omega')
[\boldsymbol{\phi}_{r}(\boldsymbol{q}',i\Omega') \cdot
\boldsymbol{\phi}_{r}(-\boldsymbol{q},-i\Omega) ] - i
\frac{1}{\beta} \sum_{i\Omega'} \sum_{\boldsymbol{q}'}
\boldsymbol{\mathcal{X}}(\boldsymbol{q}-\boldsymbol{q}',i\Omega-i\Omega')
\cdot [\boldsymbol{\phi}_{r}(\boldsymbol{q}',i\Omega') \times
\boldsymbol{\phi}_{r}(-\boldsymbol{q},-i\Omega) ] \Bigr\} . \nn
\eqa One can evaluate the four-point interaction vertex, given by
\cite{Chubukov_Review} \bqa && \lambda_{4}(\boldsymbol{q},i\Omega)
\approx \lambda_{4} \frac{|\Omega|}{|\boldsymbol{q}|^{2}} \eqa
asymptotically, which implies that this interaction is marginal in
two dimensions.

Integrating over molecular fields and performing the Fourier
transformation in Eq. (11), it is not difficult to understand the
reason why we are saying that only the center-of-mass dynamics is
considered, \bqa && {\cal S}_{eff} = \int_{0}^{\beta} d \tau \int
d^{d} \boldsymbol{r} \boldsymbol{\phi}_{r}(\boldsymbol{r},\tau)
\Bigl( -
\partial_{\tau}^{2} - v_{\phi}^{2}
\boldsymbol{\nabla}_{\boldsymbol{r}}^{2} + \gamma_{\psi}
\sqrt{-\partial_{\tau}^{2}} + \xi^{-2} \Bigr)
\boldsymbol{\phi}_{r}(\boldsymbol{r},\tau) \nn && +
\int_{0}^{\beta} d \tau_1 \int d^{d} \boldsymbol{r}_1
\int_{0}^{\beta} d \tau_2 \int d^{d} \boldsymbol{r}_2
\int_{0}^{\beta} d \tau_3 \int d^{d} \boldsymbol{r}_3 \Bigl\{
\frac{1}{\lambda_{4}(\boldsymbol{r}_1-\boldsymbol{r}_2,\tau_1-\tau_2)}
[\boldsymbol{\phi}_{r}(\boldsymbol{r}_1,\tau_1) \cdot
\boldsymbol{\phi}_{r}(\boldsymbol{r}_3,\tau_3) ]
[\boldsymbol{\phi}_{r}(\boldsymbol{r}_2,\tau_2) \cdot
\boldsymbol{\phi}_{r}(\boldsymbol{r}_4,\tau_4) ] \nn && +
\frac{1}{\lambda_{4}(\boldsymbol{r}_1-\boldsymbol{r}_2,\tau_1-\tau_2)}
[\boldsymbol{\phi}_{r}(\boldsymbol{r}_1,\tau_1) \times
\boldsymbol{\phi}_{r}(\boldsymbol{r}_3,\tau_3) ] \cdot
[\boldsymbol{\phi}_{r}(\boldsymbol{r}_2,\tau_2) \times
\boldsymbol{\phi}_{r}(\boldsymbol{r}_3,\tau_3) ] \Bigr\} . \eqa An
interesting point is that the interaction vertex appears in the
denominator instead of the numerator. Of course, this is
consistent with its dimension, considering how it rescales. This
relation reminds us of the weak-coupling and strong-coupling
duality.

\section{Beyond the Eliashberg framework}

\subsection{Renormalization for the interaction vertex in the RPA level}

An important point beyond all previous studies is to introduce
renormalization of the interaction vertex. We perform the
renormalization within the RPA analysis. It is straightforward to
see that no renormalization occurs from vector spin-chirality
fluctuations, given by \bqa && \mathcal{S}_{int}^{\mathcal{X}}
\approx \frac{1}{2} \sum_{i\Omega} \sum_{\boldsymbol{q}}
\frac{1}{\beta} \sum_{i\Omega'} \sum_{\boldsymbol{q}'}
\frac{1}{\beta} \sum_{i\Omega''} \sum_{\boldsymbol{q}''}
\frac{1}{\beta} \sum_{i\Omega'''} \sum_{\boldsymbol{q}'''}
\Bigl\langle [\boldsymbol{\phi}_{r}(\boldsymbol{q}',i\Omega')
\times \boldsymbol{\phi}_{r}(-\boldsymbol{q},-i\Omega) ]
[\boldsymbol{\phi}_{r}(\boldsymbol{q}''',i\Omega''') \times
\boldsymbol{\phi}_{r}(-\boldsymbol{q}'',-i\Omega'') ]
\Bigr\rangle_{c} \nn &&
\boldsymbol{\mathcal{X}}(\boldsymbol{q}-\boldsymbol{q}',i\Omega-i\Omega')
\boldsymbol{\mathcal{X}}(\boldsymbol{q}''-\boldsymbol{q}''',i\Omega''-i\Omega''')
= 0 , \eqa where the subscript ``c" denotes ``connected" for the
diagrammatic expansion. On the other hand, valence bond
fluctuations give rise to \bqa && \mathcal{S}_{int}^{\mathcal{V}}
\approx \frac{1}{2} \sum_{i\Omega} \sum_{\boldsymbol{q}}
\frac{1}{\beta} \sum_{i\Omega'} \sum_{\boldsymbol{q}'}
\frac{1}{\beta} \sum_{i\Omega''} \sum_{\boldsymbol{q}''}
\frac{1}{\beta} \sum_{i\Omega'''} \sum_{\boldsymbol{q}'''}
\Bigl\langle [\boldsymbol{\phi}_{r}(\boldsymbol{q}',i\Omega')
\cdot \boldsymbol{\phi}_{r}(-\boldsymbol{q},-i\Omega) ]
[\boldsymbol{\phi}_{r}(\boldsymbol{q}''',i\Omega''') \cdot
\boldsymbol{\phi}_{r}(-\boldsymbol{q}'',-i\Omega'') ]
\Bigr\rangle_{c} \nn &&
\mathcal{V}(\boldsymbol{q}-\boldsymbol{q}',i\Omega-i\Omega')
\mathcal{V}(\boldsymbol{q}''-\boldsymbol{q}''',i\Omega''-i\Omega''')
\nn && \approx \sum_{i\Omega} \sum_{\boldsymbol{q}}
\frac{1}{\beta} \sum_{i\Omega'} \sum_{\boldsymbol{q}'}
\Bigl\langle \boldsymbol{\phi}_{r}(\boldsymbol{q},i\Omega) \cdot
\boldsymbol{\phi}_{r}(-\boldsymbol{q},-i\Omega) \Bigr\rangle_{c}
\Bigl\langle \boldsymbol{\phi}_{r}(\boldsymbol{q}',i\Omega') \cdot
\boldsymbol{\phi}_{r}(-\boldsymbol{q}',-i\Omega') \Bigr\rangle_{c}
\mathcal{V}(\boldsymbol{q}-\boldsymbol{q}',i\Omega-i\Omega')
\mathcal{V}(-\boldsymbol{q}+\boldsymbol{q}',-i\Omega+i\Omega') \nn
&& = \frac{1}{4} \sum_{i\Omega} \sum_{\boldsymbol{q}} \Bigl\{
\frac{1}{\beta} \sum_{i\Omega'} \sum_{\boldsymbol{q}'}
g_{\phi}(\boldsymbol{q}+\boldsymbol{q}',i\Omega+i\Omega')
g_{\phi}(\boldsymbol{q}',i\Omega') \Bigr\}
\mathcal{V}(\boldsymbol{q},i\Omega)
\mathcal{V}(-\boldsymbol{q},-i\Omega) , \eqa where
$g_{\phi}(\boldsymbol{q},i\Omega) = \Bigl( \Omega^{2} +
v_{\phi}^{2} \boldsymbol{q}^{2} + \gamma_{\psi} |\Omega|
\Bigr)^{-1}$ is the spin-fluctuation propagator.

\subsection{Self-energy corrections to antiferromagnetic
fluctuations}

The renormalized interaction vertex causes the self-energy
correction for spin fluctuations, given by \bqa &&
\mathcal{S}_{int}^{\boldsymbol{\phi}} \approx 2 \sum_{i\Omega}
\sum_{\boldsymbol{q}} \frac{1}{\beta} \sum_{i\Omega'}
\sum_{\boldsymbol{q}'} \Bigl\langle
\mathcal{V}(\boldsymbol{q}-\boldsymbol{q}',i\Omega-i\Omega')
\mathcal{V}(-\boldsymbol{q}+\boldsymbol{q}',-i\Omega+i\Omega')
\Bigr\rangle_{c} \Bigl\langle
\boldsymbol{\phi}_{r}(\boldsymbol{q}',i\Omega') \cdot
\boldsymbol{\phi}_{r}(-\boldsymbol{q}',-i\Omega') \Bigr\rangle_{c}
\nn && \boldsymbol{\phi}_{r}(\boldsymbol{q},i\Omega) \cdot
\boldsymbol{\phi}_{r}(-\boldsymbol{q},-i\Omega) = \sum_{i\Omega}
\sum_{\boldsymbol{q}} \Bigl\{ \frac{1}{2} \frac{1}{\beta}
\sum_{i\Omega'} \sum_{\boldsymbol{q}'} G_{\mathcal{V}}
(\boldsymbol{q}-\boldsymbol{q}',i\Omega-i\Omega')
g_{\boldsymbol{\phi}}(\boldsymbol{q}',i\Omega') \Bigr\}
\boldsymbol{\phi}_{r}(\boldsymbol{q},i\Omega) \cdot
\boldsymbol{\phi}_{r}(-\boldsymbol{q},-i\Omega) , \eqa where the
RPA propagator for valence bond fluctuations is \bqa &&
G_{\mathcal{V}}(\boldsymbol{q},i\Omega) =
\frac{1}{\frac{\lambda_{4}(\boldsymbol{q},i\Omega)}{4} +
\frac{1}{4} \frac{1}{\beta} \sum_{i\Omega''}
\sum_{\boldsymbol{q}''}
g_{\phi}(\boldsymbol{q}+\boldsymbol{q}'',i\Omega+i\Omega'')
g_{\phi}(\boldsymbol{q}'',i\Omega'')} . \eqa As a result, the
effective action for spin fluctuations reads \bqa &&
\mathcal{S}_{eff}^{\boldsymbol{\phi}} = \sum_{i\Omega}
\sum_{\boldsymbol{q}}
\boldsymbol{\phi}_{r}(\boldsymbol{q},i\Omega) \Bigl[ \Omega^{2} +
v_{\phi}^{2} \boldsymbol{q}^{2} + \gamma_{\psi} |\Omega| \nn && +
\frac{1}{2} \frac{1}{\beta} \sum_{i\Omega'} \sum_{\boldsymbol{q}'}
\frac{ g_{\boldsymbol{\phi}}(\boldsymbol{q}',i\Omega')
}{\frac{\lambda_{4}(\boldsymbol{q}-\boldsymbol{q}',i\Omega-i\Omega')}{4}
+ \frac{1}{4} \frac{1}{\beta} \sum_{i\Omega''}
\sum_{\boldsymbol{q}''}
g_{\phi}(\boldsymbol{q}-\boldsymbol{q}'+\boldsymbol{q}'',i\Omega-i\Omega'+i\Omega'')
g_{\phi}(\boldsymbol{q}'',i\Omega'')} \Bigr]
\boldsymbol{\phi}_{r}(-\boldsymbol{q},-i\Omega) . \eqa This
expression shows the main point of our study. If the RPA
renormalization is not introduced, the self-energy correction for
spin fluctuations is reduced to that of the previous study
\cite{Beyond_Eliashberg_Max}. Indeed, one can find logarithmic
singularities from the ``bare" interaction. However, we find that
the self-energy correction in the RPA valence-bond fluctuation
propagator overcomes the ``bare" part. In other words, we obtain
$G_{\mathcal{V}}(\boldsymbol{q},i\Omega) \approx \bigl\{
\frac{1}{4} \frac{1}{\beta} \sum_{i\Omega''}
\sum_{\boldsymbol{q}''}
g_{\phi}(\boldsymbol{q}+\boldsymbol{q}'',i\Omega+i\Omega'')
g_{\phi}(\boldsymbol{q}'',i\Omega'') \bigr\}^{-1}$. Inserting this
expression into Eq. (18), we find \bqa &&
\mathcal{S}_{eff}^{\boldsymbol{\phi}} \approx \sum_{i\Omega}
\sum_{\boldsymbol{q}}
\boldsymbol{\phi}_{r}(\boldsymbol{q},i\Omega) \Bigl[ \gamma_{\psi}
|\Omega| + v_{\phi}^{2} \boldsymbol{q}^{2} - \frac{2
v_{\phi}^{2}}{\mathcal{C}} \frac{\gamma_{\psi}^{2}
\Omega^{2}}{v_{\phi}^{4} q^{4}} \ln (- i \Omega) \Bigr]
\boldsymbol{\phi}_{r}(-\boldsymbol{q},-i\Omega) . \eqa Detailed
calculations are shown in appendix. As shown in this expression,
$z = 3$ antiferromagnetic quantum criticality emerges in $q <
q_{c}$, where $q_{c}$ is a characteristic momentum given by the
comparison between the first term and others. This leads us to
conclude that the $z = 3$ Hertz-Moriya-Millis-Chubukov theory
describes the antiferromagnetic quantum critical point at low
temperatures.

\section{A strong-coupling approach}

\subsection{An effective field theory}

It is not easy to understand the underlying physics for the
emergence of the $z = 3$ antiferromagnetic quantum criticality
from the $z = 2$ Hertz-Moriya-Millis-Chubukov theory. Here, ``to
understand the underlying physics" means to identify elementary
excitations, resulting in $z = 3$. As described before, the
renormalization of the interaction vertex from valence bond
fluctuations is the source of the $z = 3$ quantum criticality. In
this respect our question is as follows. Can we find the dual
field variable of valence bond fluctuations? We claim that
fermionized skyrmion excitations play the role of valence bond
fluctuations. We show that fermionized skyrmion excitations with
their Fermi surface result in $z = 3$ quantum criticality.

We start from the following spin-fermion model without the kinetic
energy term of spin fluctuations, \bqa && Z =
\int{Dc_{n\sigma}D\boldsymbol{\phi}} e^{- \int_{0}^{\beta}{d\tau}
\int d^{d} \boldsymbol{r} {\cal L}} , \nn && {\cal L} =
c_{n\sigma}^{\dagger} (\partial_{\tau} - \mu_{c}) c_{n\sigma} +
\frac{1}{2m_{c}} |(\boldsymbol{\nabla} - i \boldsymbol{A})
c_{n\sigma}|^{2} + \lambda_{\phi} e^{i
\boldsymbol{Q}\cdot\boldsymbol{r}} \boldsymbol{\phi} \cdot
c_{n\alpha}^{\dagger} \boldsymbol{\sigma}_{\alpha\beta} c_{n\beta}
, \eqa regarded as the strong coupling limit of Eq. (1). We
introduce the CP$^{1}$ representation for the angular field
variable, where the amplitude field is assumed to be frozen in the
low energy limit, given by \cite{Spin_textbook} \bqa &&
\lambda_{\phi} e^{i \boldsymbol{Q}\cdot\boldsymbol{r}}
\boldsymbol{\phi} \cdot\boldsymbol{\vec{\sigma}}_{\alpha\beta} =
\lambda_{\phi} e^{i \boldsymbol{Q}\cdot\boldsymbol{r}}
|\boldsymbol{\phi}| U_{\alpha\gamma} \sigma^{z}_{\gamma\delta}
U_{\delta\beta}^{\dagger} . \eqa $\boldsymbol{U} =
\left( \begin{array}{cc} z_{\uparrow} & z_{\downarrow}^{\dagger} \\
z_{\downarrow} & - z_{\uparrow}^{\dagger} \end{array} \right)$ is
an SU(2) matrix field to represent the angular fluctuation, where
$z_{\uparrow} = e^{-i \frac{\phi}{2}} \cos \frac{\theta}{2}$ and
$z_{\downarrow} = e^{i \frac{\phi}{2}} \sin \frac{\theta}{2}$
denote complex boson fields, identified with spinons.

An essential ansatz in the strong coupling limit of
$\lambda_{\phi}$ is that the spin of an electron field is screened
from the angular fluctuation of the spin-fluctuation field,
represented as follows \bqa && \psi_{n\alpha} =
U_{\alpha\beta}^{\dagger}c_{n\beta} .\eqa $\psi_{n\alpha}$ is a
holon field which carries the same charge quantum number as an
electron while it does not carry the spin quantum number. Based on
this ansatz, we rewrite the spin-fermion model in the strong
coupling limit as follows \bqa && Z = \int D \psi_{n\alpha} D
U_{\alpha\beta} \delta(U^{\dagger}_{\alpha\gamma}U_{\gamma\beta} -
\delta_{\alpha\beta}) e^{ - \int_{0}^{\beta} d \tau \int d^{2} r
{\cal L} } , \nn && {\cal L} =
\psi_{n\alpha}^{\dagger}[(\partial_{\tau} - \mu_{c} - m_{c}^{-1})
\delta_{\alpha\beta} +
U_{\alpha\gamma}^{\dagger}\partial_{\tau}U_{\gamma\beta}]
\psi_{n\beta} + \frac{1}{2m_{c}} \Bigl(\partial_{\mathbf{r}}
\psi_{n\beta}^{\dagger} - \psi_{n\alpha}^{\dagger}
U_{\alpha\gamma}^{\dagger} (\partial_{\mathbf{r}} U_{\gamma\beta})
\Bigr) \Bigl(\partial_{\mathbf{r}} \psi_{n\beta} -
(\partial_{\mathbf{r}}U^{\dagger})_{\beta\gamma}U_{\gamma\alpha}
\psi_{n\alpha} \Bigr) \nn && + \lambda_{\phi} e^{i
\boldsymbol{Q}\cdot\boldsymbol{r}} |\boldsymbol{\phi}| \sigma
\psi_{n\sigma}^{\dagger} \psi_{n\sigma} + \frac{1}{2m_{c}}
\psi_{n\alpha}^{\dagger} \Bigl(\partial_{\mathbf{r}}U^{\dagger} +
[U^{\dagger}\partial_{\mathbf{r}}U]U^{\dagger}\Bigr)_{\alpha\gamma}
\Bigl(\partial_{\mathbf{r}}U +
U[(\partial_{\mathbf{r}}U^{\dagger})U]\Bigr)_{\gamma\beta}
\psi_{n\beta} - \frac{1}{2m_{c}} \partial_{\mathbf{r}}
(\psi_{\alpha}^{\dagger}\psi_{\alpha}) ,  \eqa where holons and
spinons appear to realize the spin-charge separation. Here, we
omit the electromagnetic vector potential for the time being. We
emphasize that this expression is just a reformulation of Eq.
(20), regarded as the change of variables.

Introducing the nonabelian gauge field of
$\mathcal{A}_{\alpha\beta}^{\nu} = - i
[\{\partial_{\nu}U^{\dagger}\}U]_{\alpha\beta}$, one can rewrite
the above as follows \bqa && Z = \int D \psi_{n\alpha} D
U_{\alpha\beta} D \mathcal{A}_{\alpha\beta}^{\nu}
\delta(U^{\dagger}_{\alpha\gamma}U_{\gamma\beta} -
\delta_{\alpha\beta}) \delta(\mathcal{A}_{\alpha\beta}^{\nu} + i
[\{\partial_{\nu}U^{\dagger}\}U]_{\alpha\beta})
\delta(\partial_{r}\mathcal{A}_{\alpha\beta}^{r}) e^{ -
\int_{0}^{\beta} d \tau \int d^{2} r {\cal L} } , \nn && {\cal L}
= \psi_{n\alpha}^{\dagger}[(\partial_{\tau} - \mu_{c} - m_{c}^{-1}
+ \lambda_{\phi} e^{i \boldsymbol{Q}\cdot\boldsymbol{r}}
|\boldsymbol{\phi}| \alpha) \delta_{\alpha\beta} - i
\mathcal{A}_{\alpha\beta}^{\tau}] \psi_{n\beta} + \frac{1}{2m_{c}}
\Bigl(\partial_{\mathbf{r}} \psi_{n\beta}^{\dagger} + i
\psi_{n\alpha}^{\dagger} \mathcal{A}_{\alpha\beta}^{\mathbf{r}}
\Bigr) \Bigl(\partial_{\mathbf{r}}\psi_{n\beta} - i
\mathcal{A}_{\beta\gamma}^{\mathbf{r}} \psi_{n\gamma} \Bigr) \nn
&& + \frac{1}{2m_{c}} \psi_{n\alpha}^{\dagger}
\Bigl(\partial_{\mathbf{r}}U^{\dagger}_{\alpha\gamma} - i
\mathcal{A}_{\alpha\delta}^{\mathbf{r}} U^{\dagger}_{\delta\gamma}
\Bigr) \Bigl(\partial_{\mathbf{r}}U_{\gamma\beta} + i
U_{\gamma\xi} \mathcal{A}_{\xi\beta}^{\mathbf{r}} \Bigr)
\psi_{n\beta} , \eqa where SU(2) gauge symmetry becomes shown
explicitly.

Although exactness is still guaranteed in Eq. (24), it is rather
complicated to handle this effective field theory. In this respect
we perform the U(1) projection, considering the subset of the
SU(2) gauge theory, given by \bqa && Z = \int D \psi_{n\sigma} D
z_{\sigma} D a_{\mu} \delta(|z_{\sigma}|^{2} - 1)
\delta(\partial_{r}a_{r}) e^{ - \int_{0}^{\beta} d \tau \int d^{2}
r {\cal L} } , ~~~~~ {\cal L} = {\cal L}_{\psi} + {\cal L}_{z} ,
\nn && {\cal L}_{\psi} = \psi_{n\sigma}^{\dagger} (
\partial_{\tau} - \mu_{r} + \lambda_{\phi} e^{i \boldsymbol{Q}\cdot\boldsymbol{r}}
|\boldsymbol{\phi}| \sigma - i \sigma a_{\tau}) \psi_{n\sigma} +
\frac{1}{2m_{c}} |(\partial_{\mathbf{r}} - i \sigma a_{\mathbf{r}}
- i A_{\mathbf{r}} ) \psi_{n\sigma}|^{2} , \nn && {\cal L}_{z} =
\frac{1}{2u_{c}} |(\partial_{\tau} - i a_{\tau} )z_{\sigma}|^{2} +
\frac{\rho_{s}}{2m_{c}} |(\partial_{\mathbf{r}} - i a_{\mathbf{r}}
)z_{\sigma}|^{2} \eqa with $\mu_{r} = \mu_{c} + m_{c}^{-1}$. One
may understand this procedure as the staggered-flux ansatz in the
SU(2) slave-boson theory \cite{SU2_SBGT_Review}, where the SU(2)
gauge symmetry is reduced to the U(1) symmetry. It is interesting
to see that the internal gauge field $a_{\mathbf{r}}$ couples to
the spin current of the renormalized electron field while the
electromagnetic field $A_{\mathbf{r}}$ does to the charge current.
The spinon part is reduced to the CP$^{1}$ representation of the
O(3) nonlinear $\sigma$ model \cite{Spin_textbook} with a
stiffness parameter of $\rho_{s} = \langle \sigma
\psi_{n\sigma}^{\dagger} \psi_{n\sigma} \rangle$, where the time
derivative term is added explicitly. This time derivative term is
expected to appear from quantum corrections, i.e., the self-energy
correction to the spinon dynamics.

\subsection{Duality transformation}

To pull out dynamics of skyrmions, we perform the duality
transformation. Unfortunately, the duality transformation for the
SU(2) case is not known. We take the easy plane approximation \bqa
&& z_{\sigma} = \frac{1}{\sqrt{2}} e^{i\phi_{\sigma}} . \eqa In
this representation the topological excitation corresponding to a
vortex is a meron instead of a skyrmion \cite{DQCP}. Since a meron
solution can be regarded as a half skyrmion, we will introduce a
meron pair as a skyrmion \cite{Nayak}.

Resorting to the easy plane approximation, we obtain \bqa && {\cal
L} = \psi_{n\sigma}^{\dagger} (
\partial_{\tau} - \mu_{r} + \lambda_{\phi} e^{i \boldsymbol{Q}\cdot\boldsymbol{r}}
|\boldsymbol{\phi}| \sigma - i \sigma a_{\tau}) \psi_{n\sigma} +
\frac{1}{2m_{c}} |(\partial_{\mathbf{r}} - i \sigma a_{\mathbf{r}}
- i A_{\mathbf{r}} ) \psi_{n\sigma}|^{2} + \frac{1}{2u_{c}}
(\partial_{\tau} \phi_{\sigma} - a_{\tau})^{2} +
\frac{\rho_{s}}{2m_{c}} |
\partial_{r} \phi_{\sigma} - a_{r} |^{2} . \nn \eqa
Performing the duality transformation, we obtain the dual
Lagrangian \bqa && {\cal L}_{D} = \psi_{n\sigma}^{\dagger} (
\partial_{\tau} - \mu_{r} + \lambda_{\phi} e^{i \boldsymbol{Q}\cdot\boldsymbol{r}}
|\boldsymbol{\phi}| \sigma - i \sigma a_{\tau}) \psi_{n\sigma} +
\frac{1}{2m_{c}} |(\partial_{\mathbf{r}} - i \sigma a_{\mathbf{r}}
- i A_{\mathbf{r}} ) \psi_{n\sigma}|^{2} \nn && + \frac{u_{c}}{2}
\bigl(\epsilon_{\tau\mu\nu}\partial_{\mu}c_{\nu}^{\sigma}
\bigr)^{2} + \frac{m_{c}}{2 \rho_{s}}
(\epsilon_{r\mu\nu}\partial_{\mu}c_{\nu}^{\sigma})^{2} - i
J_{D\mu}^{\sigma} c_{\mu}^{\sigma} - i a_{\mu}
\epsilon_{\mu\nu\lambda} \partial_{\nu} c_{\lambda}^{\sigma} ,
\eqa where $J_{D\mu}^{\sigma}$ is a meron current and
$c_{\mu}^{\sigma}$ is the dual gauge field representing spin wave
excitations.

Introducing the second quantization for meron dynamics
\cite{DQCP}, we obtain \bqa && {\cal L}_{D} =
\psi_{n\sigma}^{\dagger} (
\partial_{\tau} - \mu_{r} + \lambda_{\phi} e^{i \boldsymbol{Q}\cdot\boldsymbol{r}}
|\boldsymbol{\phi}| \sigma - i \sigma a_{\tau}) \psi_{n\sigma} +
\frac{1}{2m_{c}} |(\partial_{\mathbf{r}} - i \sigma a_{\mathbf{r}}
- i A_{\mathbf{r}} ) \psi_{n\sigma}|^{2} \nn && + |(\partial_{\mu}
- i c_{\mu}^{\sigma}) \Phi_{D}^{\sigma}|^{2} +
m_{D}^{2}|\Phi_{D}^{\sigma}|^{2} + \frac{u_{D}}{2}
(|\Phi_{D}^{\sigma}|^{2})^{2} + V_{an}[\Phi_{D}^{\sigma}] +
\frac{u_{c}}{2}
\bigl(\epsilon_{\tau\mu\nu}\partial_{\mu}c_{\nu}^{\sigma}
\bigr)^{2} + \frac{m_{c}}{2 \rho_{s}}
(\epsilon_{r\mu\nu}\partial_{\mu}c_{\nu}^{\sigma})^{2} , \eqa
where $\Phi_{D}^{\sigma}$ is a meron field and
$V_{an}[\Phi_{D}^{\sigma}]$ is an anisotropy potential for merons.

To find skyrmion dynamics, we rewrite the above meron Lagrangian
in terms of meron pairs. This derivation is far from rigorous, but
sometimes utilized in order to make physical excitations
\cite{Nayak}. Considering $\Phi_{D}^{\sigma} \longrightarrow
e^{i\theta_{\sigma}}$, the meron kinetic energy term is \bqa &&
(\partial_{\mu}\theta_{\sigma} - c_{\mu}^{\sigma})^{2} =
(\partial_{\mu}\theta_{\uparrow} - c_{\mu}^{\uparrow})^{2} +
(\partial_{\mu}\theta_{\downarrow} - c_{\mu}^{\downarrow})^{2} \nn
&& = \frac{1}{2} \Bigl([\partial_{\mu}\theta_{\uparrow} +
\partial_{\mu}\theta_{\downarrow}] - [c_{\mu}^{\uparrow} +
c_{\mu}^{\downarrow}]\Bigr)^{2} + \frac{1}{2}
\Bigl([\partial_{\mu}\theta_{\uparrow} -
\partial_{\mu}\theta_{\downarrow}] - [c_{\mu}^{\uparrow} -
c_{\mu}^{\downarrow}] \Bigr)^{2} . \eqa The spin wave term is
rewritten as follows \bqa && \frac{u_{c}}{2}
\bigl(\epsilon_{\tau\mu\nu}\partial_{\mu}c_{\nu}^{\sigma}
\bigr)^{2} + \frac{m_{c}}{2 \rho_{s}}
(\epsilon_{r\mu\nu}\partial_{\mu}c_{\nu}^{\sigma})^{2} \nn && =
\frac{u_{c}}{4} \Bigl(
\epsilon_{\tau\mu\nu}\partial_{\mu}c_{\nu}^{\uparrow} +
\epsilon_{\tau\mu\nu}\partial_{\mu}c_{\nu}^{\downarrow} \Bigr)^{2}
+ \frac{u_{c}}{4}\Bigl(
\epsilon_{\tau\mu\nu}\partial_{\mu}c_{\nu}^{\uparrow} -
\epsilon_{\tau\mu\nu}\partial_{\mu}c_{\nu}^{\downarrow} \Bigr)^{2}
\nn && + \frac{m_{c}}{4 \rho_{s}} \Bigl(
\epsilon_{r\mu\nu}\partial_{\mu}c_{\nu}^{\uparrow} +
\epsilon_{r\mu\nu}\partial_{\mu}c_{\nu}^{\downarrow} \Bigr)^{2} +
\frac{m_{c}}{4 \rho_{s}} \Bigl(
\epsilon_{r\mu\nu}\partial_{\mu}c_{\nu}^{\uparrow} -
\epsilon_{r\mu\nu}\partial_{\mu}c_{\nu}^{\downarrow} \Bigr)^{2} .
\eqa As a result, we reach \bqa && {\cal L}_{D} =
\psi_{n\sigma}^{\dagger} (
\partial_{\tau} - \mu_{r} + \lambda_{\phi} e^{i \boldsymbol{Q}\cdot\boldsymbol{r}}
|\boldsymbol{\phi}| \sigma - i \sigma a_{\tau}) \psi_{n\sigma} +
\frac{1}{2m_{c}} |(\partial_{\mathbf{r}} - i \sigma a_{\mathbf{r}}
- i A_{\mathbf{r}} ) \psi_{n\sigma}|^{2} - i a_{\mu}
\epsilon_{\mu\nu\lambda}
\partial_{\nu} c_{\lambda}^{+} \nn && +
|(\partial_{\mu} - i c_{\mu}^{+}) \Phi_{D}^{+}|^{2} +
m_{D}^{+2}|\Phi_{D}^{+}|^{2} + \frac{u_{D}^{+}}{2}
|\Phi_{D}^{+}|^{4} + \frac{u_{c}}{2}
\bigl(\epsilon_{\tau\mu\nu}\partial_{\mu}c_{\nu}^{+}\bigr)^{2} +
\frac{m_{c}}{2 \rho_{s}} (
\epsilon_{r\mu\nu}\partial_{\mu}c_{\nu}^{+} )^{2} \nn && +
|(\partial_{\mu} - i c_{\mu}^{-}) \Phi_{D}^{-}|^{2} +
m_{D}^{-2}|\Phi_{D}^{-}|^{2} + \frac{u_{D}^{-}}{2}
|\Phi_{D}^{-}|^{4} + \frac{u_{c}}{2}
(\epsilon_{\tau\mu\nu}\partial_{\mu}c_{\nu}^{-})^{2} +
\frac{m_{c}}{2 \rho_{s}} (
\epsilon_{r\mu\nu}\partial_{\mu}c_{\nu}^{-} )^{2} +
V_{an}[\Phi_{D}^{+},\Phi_{D}^{-}] , \eqa where \bqa &&
\theta_{\pm} = \frac{\theta_{\uparrow} \pm
\theta_{\downarrow}}{\sqrt{2}} \longrightarrow \Phi_{D}^{\pm} ,
~~~~~ c_{\mu}^{\pm} = \frac{c_{\mu}^{\uparrow} \pm
c_{\mu}^{\downarrow}}{\sqrt{2}} .  \eqa

It is interesting to see that the $\Phi_{D}^{-}$ field does not
couple to dynamics of renormalized electrons $\psi_{\sigma}$
directly. Only the $\Phi_{D}^{+}$ field couples to fermions via
the mutual Chern-Simons term. Integrating over $\Phi_{D}^{+}$,
$c_{\mu}^{+}$, and $a_{\mu}$ fields, we will obtain an effective
field theory for renormalized electrons interacting with
skyrmions. It is natural to expect the emergence of nonlocal
interactions between spin currents of renormalized electrons
(holons) and skyrmion currents. Integrating over holon
excitations, we reach a dual filed theory for skyrmion dynamics.
Unfortunately, this procedure cannot be performed rigorously. We
propose the following skyrmion field theory \bqa &&
{\cal S}_{sk} = \int_{0}^{\beta_r} d \tau \int d^{2}
\boldsymbol{r} \Bigl\{ \mu_{sk} \Phi_{s}^{\dagger}
(\partial_{\tau} - i c_{\tau}) \Phi_{s} + |(\partial_{\mu} - i
c_{\mu} ) \Phi_{s}|^{2} + m_{s}^{2}|\Phi_{s}|^{2} +
\frac{u_{s}}{2} |\Phi_{s}|^{4} + \frac{u_{c}}{2}
(\epsilon_{\tau\mu\nu}\partial_{\mu}c_{\nu} )^{2} + \frac{m_{c}}{2
\rho_{s}} ( \epsilon_{r\mu\nu}\partial_{\mu}c_{\nu} )^{2} \Bigr\}
, \nn \eqa where $\Phi_{D}^{-}$ is replaced with $\Phi_{s}$ and
the $-$ superscript in $c_{\mu}^{-}$ is omitted.

It is our proposal to introduce the linear-time derivative term
into Eq. (34). Then, the linear-time derivative term becomes more
relevant than the second-time derivative term at low energies.
Unfortunately, we cannot prove the emergence of the Galileian
invariance (the linear-time derivative) from the relativistic
invariance (the second-time derivative) at present. In this
respect one can call the emergence of the Galileian invariance in
the presence of itinerant electrons as our speculation.

There must be an underlying physical mechanism for this
linear-time derivative term, which breaks the particle-hole
symmetry for skyrmion excitations. It has been demonstrated that
the density of skyrmion excitations is finite at an
antiferromagnetic quantum critical point without itinerant
electrons, i.e. in an insulating antiferromagnet
\cite{Skyrmion_RG}. Of course, only the second-time derivative
term is allowed in this case, i.e., particle-hole symmetric, which
means that an equal number of skyrmion and anti-skyrmion
excitations exists at this quantum critical point. Our problem is
what happens on the particle-hole symmetry in skyrmion excitations
at the quantum critical point where itinerant charge carriers are
introduced. This is a long-standing problem, where
non-perturbative effects from interactions between itinerant
electrons and ``many" topologically nontrivial excitations should
be taken into account on equal footing. Frankly speaking, we do
not have any reliable mathematical tools for the description of
such interactions.

Our speculation is that the presence of itinerant electrons will
induce the particle-hole symmetry breaking in the skyrmion sector
because itinerant electrons favor skyrmion excitations (or
anti-skyrmions, i.e., one of the two). The physical mechanism is
as follows. When itinerant electrons move in the background of
skyrmions, they feel an effective magnetic flux, which quenches
the kinetic energy of electrons. Our expectation is that the gain
in the electron kinetic energy contribution can overcome the
energy cost due to the particle-hole symmetry breaking in the
skyrmion sector. As a result, the particle-hole symmetry breaking
is favorable in the respect of the total energy. Actually, this
mechanism has been realized in the system of frustrated magnets,
where the presence of itinerant electrons leads the co-planar
ordering in the triangular antiferromagnet to be ordered into an
out-of-plane way, which corresponds to a spin chiral order
\cite{Kondo_Frustration}.

Possibility of statistical transmutation for vortices has been
discussed in the vortex liquid phase \cite{Fisher_Fermionization}
and at quantum criticality in geometrically frustrated spin
systems \cite{Fermionization_Vortex}. Fermionization of skyrmion
excitations can be performed in the same way as that of vortices.
The key point is that the Chern-Simons term becomes irrelevant at
quantum criticality
\cite{Fisher_Fermionization,Fermionization_Vortex}, implying that
the quantum statistics for skyrmions or vortices may not be well
defined at quantum criticality. Actually, such excitations are
strongly interacting at criticality, where we are not allowed to
pin down elementary excitations clearly. We reach an effective
field theory of fermionic skyrmions, \bqa &&
{\cal S}_{eff} = \int_{0}^{\beta_r} d \tau \int d^{2}
\boldsymbol{r} \Bigl\{ \psi_{s}^{\dagger} (\partial_{\tau} -
\mu_{\psi} - i c_{\tau}) \psi_{s} + \frac{1}{2m_{sk}}
|(\boldsymbol{\nabla} - i \boldsymbol{c} ) \psi_{s}|^{2} +
\frac{u_{c}}{2} (\epsilon_{\tau\mu\nu}\partial_{\mu}c_{\nu} )^{2}
+ \frac{m_{c}}{2 \rho_{s}} (
\epsilon_{r\mu\nu}\partial_{\mu}c_{\nu} )^{2} \Bigr\} , \eqa where
$\psi_{s}$ represents the fermionic skyrmion field with the
skyrmion chemical potential $\mu_{\psi}$ and the band mass $m_{sk}
\propto \mu_{sk}$. We would like to emphasize that Eq. (35) can be
derived from Eq. (34) at least formally via the Chern-Simons
transformation \cite{Fisher_Fermionization,Fermionization_Vortex}.

Another important point with the particle-hole symmetry breaking
term is that the skyrmion chemical potential is assumed to be
finite. As discussed before, the skyrmion density has been shown
to be finite in the O(3) nonlinear $\sigma$ model without
itinerant electrons \cite{Skyrmion_RG}. It is natural to expect
that this will hold even in the presence of itinerant electrons.
Then, Landau damping for gauge fluctuations emerges from
particle-hole excitations of fermionic skyrmions near their Fermi
surface, resulting in the $z = 3$ dynamics for spin fluctuations.

Although the $z = 3$ antiferromagnetic quantum criticality appears
in the fermionized skyrmion ansatz with the skyrmion Fermi
surface, it is still not clear how this $z = 3$ quantum
criticality can be connected with that of the previous
diagrammatic approach. It has been shown that the skyrmion field
has the same symmetry as the valence bond field in an insulating
antiferromagnet \cite{DQCP,Hermele_ASL,Wen_ASL}. Resorting to the
fermion representation for the spin operator, one can construct an
effective Dirac theory with an enhanced emergent symmetry,
compared with the Heisenberg model with the O(3) spin symmetry
\cite{Hermele_ASL,Wen_ASL,Tanaka_SO5}. Based on the Dirac theory,
one can construct the valence bond operator and skyrmion operator
in terms of the Dirac spinor. Since the symmetry property is well
defined for Dirac fermions, it is straightforward to investigate
symmetries of both valence bond and skyrmion operators. They turn
out to be the same as each other. One can perform the same work in
the boson representation for spin, where the role of the spin
Berry phase is crucial to assign the valence bond quantum number
to a skyrmion \cite{DQCP}. On the other hand, it is not verified
yet whether skyrmion excitations can be identified with valence
bond fluctuations in the presence of itinerant electrons. In
particular, the role of Berry phase is not clarified yet. However,
it is expected that valence bond fluctuations will be deeply
related with skyrmion fluctuations. Actually, the skyrmion
operator has been constructed in terms of fermion bilinear
operators, identified with the valence bond operator in itinerant
antiferromagnets \cite{Xu_Berry}.


\section{Discussion and summary}

In summary, we showed that the dynamical critical exponent can
change from $z = 2$ to $z = 3$ at low temperatures in
antiferromagnetic quantum criticality, where nonlocal interactions
between spin fluctuations, which result from Fermi surface
fluctuations, are responsible for the change of the dynamics of
spin-fluctuations. In particular, we claimed that renormalization
in the nonlocal interaction vertex should be introduced because
the renormalized part turns out to be larger than the bare
(unrenormalized) interaction. We performed the renormalization in
the RPA level, where valence bond fluctuations are introduced
explicitly and the nonlocal interaction vertex is identified with
a bare propagator of valence bond fluctuations. As a result, the
RPA renormalization in valence bond fluctuations gives rise to the
$z = 3$ antiferromagnetic quantum criticality.

We tried to understand the underlying physics for the emergence of
$z = 3$ antiferromagnetic quantum criticality. Hinted from the
fact that valence bond fluctuations can be identified with
skyrmion excitations, we constructed a dual field theory in terms
of skyrmion excitations. The key approximation is to introduce the
particle-hole symmetry breaking term, i.e., the linear-time
derivative term instead of the second-order time-derivative term.
This construction is based on the fact that the interplay between
itinerant electrons and skyrmion excitations may favor skyrmion
excitations more instead of anti-skyrmion fluctuations. Allowing
the statistical transmutation for skyrmion fluctuations, where the
Chern-Simons term turns out to be irrelevant at the quantum
critical point, we found a dual theory in terms of fermionized
skyrmion excitations interacting with spin-wave fluctuations. If
the density of skyrmions is finite at the quantum critical point,
fermionized skyrmions will form their Fermi surface. As a result,
Landau damping dynamics for spin-wave excitations is introduced to
cause the $z = 3$ antiferromagnetic quantum criticality.

We propose that this $z = 2$ to $z = 3$ crossover in two
dimensions may explain the crossover behavior in the specific heat
measurement of YbRh$_{2}$Si$_{2}$, where the logarithmic divergent
behavior of the specific heat coefficient turns into the power-law
divergent behavior of $T^{-2/3}$ at low temperatures
\cite{Specific_Heat1,Specific_Heat2}.

An important missing point in our study is to neglect higher-order
terms beyond the fourth order expansion. Such higher-order terms
have been argued to be important because the two-dimensional $z =
2$ spin-fluctuation theory turns out to allow infinite number of
marginal interactions, which originate from singular momentum and
frequency dependencies for interaction vertices. Based on this
observation, it was claimed that an anomalous dimension for spin
fluctuations appears beyond the Eliashberg framework
\cite{Chubukov_Infinitely_Marginal}. We do not exclude this
possibility. Actually, the spin susceptibility can be
$\chi_{r}(\boldsymbol{q},i\Omega) = \Bigl( \gamma_{\psi} |\Omega|
+ v_{\phi}^{2} \boldsymbol{q}^{2} - \frac{2
v_{\phi}^{2}}{\mathcal{C}} \frac{\gamma_{\psi}^{2}
\Omega^{2}}{v_{\phi}^{4} q^{4}} \ln (- i \Omega)
\Bigr)^{-(1+\gamma)}$ instead of $\chi_{r}(\boldsymbol{q},i\Omega)
= \Bigl( \gamma_{\psi} |\Omega| + v_{\phi}^{2} \boldsymbol{q}^{2}
- \frac{2 v_{\phi}^{2}}{\mathcal{C}} \frac{\gamma_{\psi}^{2}
\Omega^{2}}{v_{\phi}^{4} q^{4}} \ln (- i \Omega) \Bigr)^{-1}$ in
Eq. (19), where $\gamma$ is the anomalous dimension of spin
fluctuations. We leave the study beyond the fourth order as a
future work.

\section*{Acknowledgement}

KS was supported by the National Research Foundation of Korea
(NRF) grant funded by the Korea government (MEST) (No.
2012000550).

\appendix*

\section{Self-energy corrections to antiferromagnetic fluctuations}

The RPA self-energy correction for valence bond fluctuations is
given by \bqa && \frac{1}{\beta} \sum_{i\Omega''}
\sum_{\boldsymbol{q}''}
g_{\phi}(\boldsymbol{q}-\boldsymbol{q}'+\boldsymbol{q}'',i\Omega-i\Omega'+i\Omega'')
g_{\phi}(\boldsymbol{q}'',i\Omega'') \nn && = \frac{1}{\beta}
\sum_{i\Omega''} \sum_{\boldsymbol{q}''}
\frac{1}{(\Omega-\Omega'+\Omega'')^{2} + v_{r}^{\phi 2}
(\boldsymbol{q}-\boldsymbol{q}'+\boldsymbol{q}'')^{2} +
\gamma_{\psi} |\Omega-\Omega'+\Omega''|} \frac{1}{\Omega^{''2} +
v_{\phi}^{2} q^{''2} + \gamma_{\psi} |\Omega''|} \nn && \approx
\frac{1}{(2\pi)^{2}} \frac{1}{\beta} \sum_{i\Omega''}
\int_{0}^{\infty} d q'' q'' \int_{0}^{2\pi} d \theta
\frac{1}{v_{\phi}^{2} q^{''2} + 2 v_{\phi}^{2} q''
|\boldsymbol{q}-\boldsymbol{q}'| \cos \theta + v_{\phi}^{2}
(\boldsymbol{q}-\boldsymbol{q}')^{2} + \gamma_{\psi}
|\Omega-\Omega'+\Omega''|} \frac{1}{v_{\phi}^{2} q^{''2} +
\gamma_{\psi} |\Omega''|} \nn && \approx \frac{\mathcal{C}}{4 \pi
\gamma_{\psi} v_{\phi}^{2}} \ln
\frac{\Lambda_{\Omega}}{\Omega-\Omega'} , \eqa where we resorted
to $\boldsymbol{q} \approx \boldsymbol{q}'$ in the last line.
$\mathcal{C}$ is a positive numerical constant. This renormalized
contribution turns out to be much larger than the bare vertex of
$\lambda_{4}(\boldsymbol{q},i\Omega)$ because
$\frac{|\Omega|}{|\boldsymbol{q}|^{2}} \ll 1$ should be satisfied
in the expansion. As a result, the self-energy correction in spin
fluctuations is given by \bqa && \frac{1}{\beta} \sum_{i\Omega'}
\sum_{\boldsymbol{q}'} \frac{
g_{\boldsymbol{\phi}}(\boldsymbol{q}',i\Omega')
}{\frac{\lambda_{4}(\boldsymbol{q}-\boldsymbol{q}',i\Omega-i\Omega')}{4}
+ \frac{1}{\beta} \sum_{i\Omega''} \sum_{\boldsymbol{q}''}
g_{\phi}(\boldsymbol{q}-\boldsymbol{q}'+\boldsymbol{q}'',i\Omega-i\Omega'+i\Omega'')
g_{\phi}(\boldsymbol{q}'',i\Omega'')} \nn && \approx
\frac{1}{\beta} \sum_{i\Omega'} \sum_{\boldsymbol{q}'} \frac{
g_{\boldsymbol{\phi}}(\boldsymbol{q}',i\Omega') }{ \frac{1}{\beta}
\sum_{i\Omega''} \sum_{\boldsymbol{q}''}
g_{\phi}(\boldsymbol{q}-\boldsymbol{q}'+\boldsymbol{q}'',i\Omega-i\Omega'+i\Omega'')
g_{\phi}(\boldsymbol{q}'',i\Omega'')} \nn && \approx
\frac{1}{\beta} \sum_{i\Omega'} \frac{
g_{\boldsymbol{\phi}}(\boldsymbol{q},i\Omega') }{ \frac{1}{\beta}
\sum_{i\Omega''} \sum_{\boldsymbol{q}''}
g_{\phi}(\boldsymbol{q}'',i\Omega-i\Omega'+i\Omega'')
g_{\phi}(\boldsymbol{q}'',i\Omega'')} \approx \frac{4 \pi
\gamma_{\psi} v_{\phi}^{2}}{\mathcal{C}} \frac{1}{\beta}
\sum_{i\Omega'} \frac{1}{\ln \Bigl(
\frac{\Lambda_{\Omega}}{\Omega-\Omega'} \Bigr)} \frac{1}{
v_{\phi}^{2} q^{2} + \gamma_{\psi} |\Omega'|}
\nn && = \frac{4 \gamma_{r}^{\psi} v_{r}^{\phi 2}}{\pi\mathcal{C}}
\int_{0}^{\infty} d \nu \ln(\nu - i\Omega) \frac{\gamma_{\psi}
\nu}{\gamma_{\psi}^{2} \nu^{2} + v_{\phi}^{4} q^{4}} \nn && =
\frac{2 v_{r}^{\phi 2}}{\mathcal{C}} \Bigl\{ \ln (- i \Omega) \ln
\Bigl( \frac{- \gamma_{\psi}^{2} \Omega^{2} + v_{\phi}^{4}
q^{4}}{v_{\phi}^{4} q^{4}} \Bigr) -
PolyLog\Bigl(2,\frac{\gamma_{\psi} \Omega}{\gamma_{\psi} \Omega -
v_{\phi}^{2} q^{2}}\Bigr) - PolyLog\Bigl(2,\frac{\gamma_{\psi}
\Omega}{\gamma_{\psi} \Omega + v_{\phi}^{2} q^{2}}\Bigr) \Bigr\} .
\eqa Considering that the $PolyLog$ function is not singular at
low momenta and frequencies, we expand the first $log$ term in
$|\Omega|/ |\boldsymbol{q}|^{2} \ll 1$ and find the expression of
Eq. (19).


\begin{thebibliography}{9}
\bibitem{Ising_Duality_Review} J. B. Kogut, Rev. Mod. Phys. {\bf
51}, 659 (1979).
\bibitem{Tesanovic_vs_Kleinert} F. S. Nogueira and H. Kleinert,
arXiv:cond-mat/0303485, to appear in the World Scientific review
volume "Order, Disorder, and Criticality", Edited by Y. Holovatch.
\bibitem{Bosonization_textbook} A. O. Gogolin, A. A. Nersesyan, and A.
M. Tsvelik, \textit{Bosonization and Strongly Correlated Systems}
(Cambridge University Press, New York, 2004).
\bibitem{Pepin_Chubukov} J. Rech, C. Pepin, and A. V. Chubukov, Phys. Rev. B
{\bf 74}, 195126 (2006).
\bibitem{Beyond_Eliashberg_Lee} S.-S. Lee, Phys. Rev. B {\bf 80}, 165102
(2009).
\bibitem{Beyond_Eliashberg_Max} M. A. Metlitski and S. Sachdev, Phys. Rev. B
{\bf 82}, 075128 (2010).
\bibitem{Chubukov_Review} Ar. Abanov, A. V. Chubukov, and J. Schmalian,
Adv. Phys. {\bf 52}, 119 (2003).
\bibitem{Spin_textbook} A. Auerbach, \textit{Interacting Electrons and Quantum
magnetism }(Springer-Verlag, New York, 1994).
\bibitem{SU2_SBGT_Review} P. A. Lee, N. Nagaosa, and X.-G. Wen, Rev. Mod. Phys.
{\bf 78}, 17 (2006).
\bibitem{DQCP} T. Senthil, A. Vishwanath, L. Balents, S. Sachdev, and
M. P. A. Fisher, Science {\bf 303}, 1490 (2004); T. Senthil, L.
Balents, S. Sachdev, A. Vishwanath, and M. P. A. Fisher, Phys.
Rev. B {\bf 70}, 144407 (2004).
\bibitem{Nayak} C. Nayak, Phys. Rev. Lett. {\bf 85}, 178 (2000).
\bibitem{Skyrmion_RG} Z. Nazario and D. I. Santiago, Phys. Rev.
Lett. {\bf 97}, 197201 (2006); Z. Nazario and D. I. Santiago,
arXiv:cond-mat/0611383.
\bibitem{Kondo_Frustration} I. Martin and C. D. Batista,
Phys. Rev. Lett. {\bf 101}, 156402 (2008); Y. Kato, I. Martin, and
C. D. Batista, Phys. Rev. Lett. {\bf 105}, 266405 (2010); G.-W.
Chern, Phys. Rev. Lett. {\bf 105}, 226403 (2010); Y. Akagi and Y.
Motome, J. Phys. Soc. Jpn. {\bf 79}, 083711 (2010); Y. Akagi, M.
Udagawa, and Y. Motome, Phys. Rev. Lett. {\bf 108}, 096401 (2012).
\bibitem{Fisher_Fermionization} V. M. Galitski, G. Refael, M. P.
A. Fisher, and T. Senthil, Phys. Rev. Lett. {\bf 95}, 077002
(2005).
\bibitem{Fermionization_Vortex} J. Alicea, O. I. Motrunich, and M.
P. A. Fisher, Phys. Rev. Lett. {\bf 95}, 247203 (2005); J. Alicea,
O. I. Motrunich, and M. P. A. Fisher, Phys. Rev. B {\bf 73},
174430 (2006); S. Ryu, O. I. Motrunich, J. Alicea, and M. P. A.
Fisher, Phys. Rev. B {\bf 75}, 184406 (2007).
\bibitem{Hermele_ASL} M. Hermele, T. Senthil, and M. P. A. Fisher,
Phys. Rev. B {\bf 72}, 104404 (2005).
\bibitem{Wen_ASL} Y. Ran and X.-G. Wen,
arXiv:cond-mat/0609620v3.
\bibitem{Tanaka_SO5} A. Tanaka and X. Hu, Phys. Rev. Lett. {\bf
95}, 036402 (2005).
\bibitem{Xu_Berry} L. Fu, S. Sachdev, and C. Xu, Phys. Rev. B
{\bf 83}, 165123 (2011).
\bibitem{Specific_Heat1} R. Kuchler, N. Oeschler, P. Gegenwart,
T. Cichorek, K. Neumaier, O. Tegus, C. Geibel, J. A. Mydosh, F.
Steglich, L. Zhu, and Q. Si, Phys. Rev. Lett. {\bf 91}, 066405
(2003).
\bibitem{Specific_Heat2} K. Ishida, K. Okamoto, Y. Kawasaki, Y. Kitaoka,
O. Trovarelli, C. Geibel, and F. Steglich, Phys. Rev. Lett. {\bf
89}, 107202 (2002).
\bibitem{Chubukov_Infinitely_Marginal} Ar. Abanov and A. Chubukov,
Phys. Rev. Lett. {\bf 93}, 255702 (2004).
\end{thebibliography}
\end{document}